\documentclass[conference]{IEEEtran}
\IEEEoverridecommandlockouts
\usepackage{cite}
\usepackage{amsmath,amssymb,amsfonts}
\usepackage{graphicx}
\usepackage{textcomp}
\usepackage{xcolor}
\usepackage{algorithm}
\usepackage{tabularx}
\usepackage{float}
\usepackage{cleveref}
\usepackage[noend]{algpseudocode}

\def\BibTeX{{\rm B\kern-.05em{\sc i\kern-.025em b}\kern-.08em
    T\kern-.1667em\lower.7ex\hbox{E}\kern-.125emX}}
\begin{document}
\makeatletter
\newcommand{\newlineauthors}{%
  \end{@IEEEauthorhalign}\hfill\mbox{}\par
  \mbox{}\hfill\begin{@IEEEauthorhalign}
}
\makeatother
\title{Distributed Grid restoration based on graph theory}

\author{
\IEEEauthorblockN{Ayush Sinha}
\IEEEauthorblockA{\textit{pro.ayush@iiita.ac.in} \\
\textit{IIIT Allahabad, India}}
\and
\IEEEauthorblockN{Sourin Chakrabarti}
\IEEEauthorblockA{\textit{iit2016513@iiita.ac.in} \\
\textit{IIIT Allahabad, India}}
\and
\IEEEauthorblockN{Prof. O.P. Vyas}
\IEEEauthorblockA{\textit{opvyas@iiita.ac.in} \\
\textit{IIIT Allahabad, India}}
}
\IEEEoverridecommandlockouts
\IEEEpubid{\makebox[\columnwidth]{978-1-5386-5541-2/18/\$31.00~\copyright2018 IEEE \hfill} \hspace{\columnsep}\makebox[\columnwidth]{ }}
\maketitle

\begin{abstract}
With the emergence of smart grids as the primary means of distribution across wide areas, the importance of improving its resilience to faults and mishaps is increasing. The reliability of a distribution system depends upon its tolerance to attacks and the efficiency of restoration after an attack occurs. This paper proposes a unique approach to the restoration of smart grids under attack by impostors or due to natural calamities via optimal islanding of the grid with primary generators and distributed generators(DGs) into sub-grids minimizing the amount of load shed which needs to be incurred and at the same time minimizing the number of switching operations via graph theory. The minimum load which needs to be shed is computed in the first stage followed by selecting the nodes whose load needs to be shed to achieve such a configuration and then finally deriving the sequence of switching operations required to achieve the configuration. The proposed method is tested against standard IEEE 37-bus and a 1069-bus grid system and the minimum load shed along with the sequencing steps to optimal configuration and time to achieve such a configuration are presented which demonstrates the effectiveness of the method when compared to the existing methods in the field. Moreover, the proposed algorithm can be easily modified to incorporate any other constraints which might arise due to any operational configuration of the grid. 
\end{abstract}


\section{Introduction}
Smart grids have been emerging readily since their inception due to their capability of imbibing powers of communication and computation in electrical grids which leads to better control and management of the grids. Smart grids being fully configurable with respect to their transmission and distribution have enabled the application of various methodologies to develop better resilience to attacks or natural disasters \cite{b5}. Moreover, the problem of automatic restoration of smart grids after an anomalous event has gained importance since maximum load restoration needs to be done at the earliest to minimize losses to the economy.

 Typically, the reliability of a grid is assured via N-1 or N-k security criterion but for resilience, various operational measures need to be employed. Hardening of smart grids is another step that could be taken to reduce the impact of attacks on the grid \cite{b1}. Grid restoration post-attack is mainly done by islanding the grid into microgrids which are self-sufficient through Distributed Generators(DGs) if not connected to the main grid itself. The reconfiguration is mainly done through switches that enable connecting tie lines and opening a normally closed bus to redistribute the load. Redistribution enables restoration of maximum possible load by using DGs in combination with the primary generator or even with other DGs.
 
 This paper primarily involves a new approach to power restoration in smart grids post attacks. The objectives are two-fold. First, the minimum amount of load which needs to be shed is calculated. Then, the sequence of switching operations required for achieving such a configuration with minimum load shed is deduced. For solving the above problems, a graph-based approach is used where the complete grid is mapped to a graph with graph nodes as the grid nodes and the edges as the lines in the grid. In the case of an attack, the graph is decomposed into several connected components. For the first stage, to find the minimum load shed which could be achieved, all the available tie lines are connected supplying as much load as possible to the areas blacked out. Once this configuration is obtained, the second stage is carried out by identifying the unnecessary tie lines which are connected. This is identified by removing the tie lines sequentially and re-computing the load shed at each of the configurations which satisfy the grid criterion. This method is quite efficient since the order in which the tie lines are disconnected does not matter which results in a very competitive complexity once combined with memoization. Once the best configuration is obtained, a multi-source search is carried out with the leaf nodes of an island as the sources. The one with the lowest impact is selected to be shed and the corresponding sectionalization switches are opened. Moreover, any such line which causes the radial topology to be disrupted such as forming a cycle is disconnected.

Finally, the approach is tested with standard IEEE 37-node and a 1069 node grid against various attacks and with various hardening plans. The minimum load shed, the optimum reconfiguration, and the time to achieve such a result is noted. Finally, it is compared to the modern state of the art techniques.

\IEEEpubidadjcol
\section{Related Work}
Smart grid resilience can primarily be enforced by either hardening the transmission buses through physical measures \cite{b2} or through various operational measures such as reconfiguration and DG islanding \cite{b3}. Reconfiguration in power systems has been particularly helpful since it enables using various energy sources in harmony and the process reduces load shed \cite{b4}. Moreover, specific importance can be given to nodes that are pivotal to the system \cite{b6}. Islanding on the other hand helps mitigate the effects of disasters by clustering the original network into smaller networks \cite{b7}. Several approaches have been researched to enable restoration of power distribution during disasters using microgrids and DG islanding such as in \cite{b8,b9}.

For resilience against attacks or disasters, hardening serves as the first layer of protection for smart grids. \cite{b11} proposed a greedy algorithm that considers the bus with a maximum load at each step to develop a corresponding hardening plan. The Defender Attacker Defender(DAD) model \cite{b12} is a widely used infrastructure for the analysis of smart grids and devising hardening plans. The DAD model iteratively finds the optimal hardening plan for a grid considering reinforced attacks at each stage. Problems formulated by the DAD model have often been solved by the Column Constraint Generation(CCG) method of solving optimization problems \cite{b13}. The DAD model was further modified to use line switching as restoration measures \cite{b14}. Moreover, the model was further developed to include certain operational constraints which would account for real-life limitations \cite{b15}. Accounting for DGs, microgrids, etc. in the DAD model to create a more resilient solution was first experimented with by \cite{b16}. \cite{b6} proposed a three-stage hardening and restoration plan based on the DAD model. It provides an optimal DG islanding strategy and a hardening plan by iteratively using the results of the previous iteration. 

Restoration of power in smart grids is often formulated as a Distributed System Restoration(DSR) problem which is primarily characterized by optimization problems with complex objective functions and several topological and operational constraints. The various techniques employed to solve these problems include expert systems \cite{b17}, artificial intelligence such as genetic algorithms \cite{b18} and fuzzy logic \cite{b19}, dynamic programming \cite{b20}, various mathematical optimization techniques \cite{b21} and several graph-based heuristic search algorithms \cite{b22}. Most of these algorithms do not account for the inclusion of DGs or microgrids in the computation of the optimum solution. These factors include various uncertainties such as bi-directional power flows, islanded configurations, etc. These factors have been tackled by \cite{b23} which account for DGs and develop a restoration plan in a two-stage process specialized for Cold Load Pick Up(CLPU) conditions and in \cite{b24} which used spanning-tree search to develop a restoration plan considering microgrids inside the primary grid. Resilient system restoration procedures have also been proposed by \cite{b25} which extends it to the case of natural disasters where several buses are to be affected at once. For time-critical computations, an approach based on sequential optimization is proposed in \cite{b28} which provides an approximately accurate solution in a much shorter time.

Smart grids can be easily modeled in the form of graphs with the feeders as graph edges between two nodes. \cite{b26} proposed a method of restoration based on minimum spanning forests. The complete grid is divided into several smaller Self-Sustained Islanded Grids(SSIGs) which operate in harmony with each other either through the main turbine or through DGs. \cite{b24} on the other hand represented the grid a spanning tree and then used selective cyclic interchange policies to generate a new spanning tree iteratively and selecting the most optimal one. \cite{b27} again used a minimum spanning tree search to propose a new algorithm for finding the optimal configuration accounting for priority information of nodes. \cite{b10} proposes a time-efficient self-healing strategy to connect the blacked-out sections of the grid to a microgrid with a DG with a rolling-horizon optimization technique. \cite{b29} devises a power restoration strategy in a radial network with the help of sectionalization switches based on Edmond’s Maximal Spanning Tree Algorithm. \cite{b31} proposed a restoration algorithm based on Dijkstra's algorithm for computation of shortest paths and used the results of a stage to reinforce the results of the next stage. \cite{b32} gives another algorithm using Kruskal's algorithm. \cite{b33} gives an analysis of the various restoration techniques and compares them on several bases. \cite{b35} provides an algorithm which could find all possible solutions to connect energized healthy zones with blacked-out zones using graph theory. It then rejects the solutions which do not satisfy all the constraints. But the algorithm doesn't account for cases where a single node from the inactive sub-graph is connected to multiple nodes from the healthy active sub-graph. Moreover, it doesn't account for DGs during load restoration.

\section{Proposed methodology}
Our proposed methodology for power restoration in smart grids can be primarily divided into certain sections. The problem is defining or choosing a mathematical formulation based on which we base our further architecture. Then, we need to devise an algorithm for restoration based on the constraints of the mathematical formulation. Devising an algorithm involves finding the minimum load shed possible along with the optimum switching sequence for reaching a grid configuration which provides the minimum load shed.
\section*{Nomenclature}
\addcontentsline{toc}{section}{Nomenclature}
\begin{IEEEdescription}[\IEEEusemathlabelsep\IEEEsetlabelwidth{$V_1,V_2,V_3$}]
\item[$i,j$] Sample bus indexes.
\item[$E$] Set of branches in the grid.
\item[$B$] Set of buses in the grid.
\item[$r$] Resistance of the feeders.
\item[$x$] Reactance of the feeders.
\item[$\varpi$] Load weight.
\item[$M$] Large number.
\item[$P_{L}$] Active Load demand.
\item[$Q_{L}$] Reactive Load demand.
\item[$\pi(j)$] Set of parent buses of j.
\item[$\delta(j)$] Set of child buses of j.
\item[$q_{ij}$] Binary variable to determine whether feeder $ij$ is operational or not.
\item[$P_{shed}$] Active load to be shed.
\item[$Q_{shed}$] Reactive Load to be shed.
\item[$P_{impact}$] Impact of load shed.
\item[$P_{DG}$] Active Load supply.
\item[$Q_{DG}$] Reactive Load supply.
\item[$H$] Active power flow.
\item[$G$] Reactive power flow.
\item[$U$] Bus voltage.
\end{IEEEdescription}
\subsection{Mathematical formulation of constraints}
An efficient restoration strategy should account for most of the topological and operational constraints which might occur in real-life implementations of the system. In a modern smart grid, it is assumed that the DGs can operate in harmony with each other and also with the primary generator. Moreover, each DG should possess the capability to operate independently in islanded situations. One of the most widely used models for the formulation of constraints for solving hardening and restorations problems is the linearized DistFlow model \cite{b30}. It has proved to be quite efficient in the representation of these kinds of problems as suggested by \cite{b13} and \cite{b6}.
The constraints according to the linearized DistFlow model can be formulated as \cref{eq:1,eq:2,eq:3,eq:4,eq:5}. Each constraint corresponds to a limitation of the power grid. More specifically, the first constraint refers to the power flow balance in the grid. The second equation corresponds to the actual DistFlow equation which states that in a closed branch the flow should be limited whereas in an open branch it should be zero. The bounds for the amount of load that can be shed at each bus is given by \cref{eq:3}. The generators are constrained by \cref{eq:4}. The voltage constraints at each bus are represented by \cref{eq:5}. 

\begin{flalign}
& \begin{cases}
P_{DG,j} - (P_{L,j}-P_{shed,j}) \\ 
\qquad \qquad = \sum\limits_{s\in \delta(j)}H_{js} - \sum\limits_{i\in \pi(j)}H_{ij}\\
Q_{DG,j} - (Q_{L,j}-Q_{shed,j}) \\
\qquad \qquad= \sum\limits_{s\in \delta(j)}G_{js} - \sum\limits_{i\in \pi(j)}G_{ij}
\end{cases},\forall j \in B &&
\label{eq:1}
\end{flalign}
\begin{flalign}
& \begin{cases}
-M(1-q_{ij}) \leq U_{j} - U_{i} \\ 
\qquad - (r_{ij}H_{ij} + x_{ij}G_{ij})/U_{0} \leq M(1-q_{ij}) \\
H_{ij},G_{ij} \in [-S_{ij}^{max}q_{ij},S_{ij}^{max}q_{ij}]
\end{cases}, \forall (i,j) \in E &&
\label{eq:2}
\end{flalign}
\begin{flalign}
& \begin{cases}
P_{shed,j} \in [0,P_{L,j}]\\
Q_{shed,j} \in [0,Q_{L,j}]
\end{cases}, \forall j \in B &&
\label{eq:3}
\end{flalign}
\begin{flalign}
& \begin{cases}
P_{DG,j} \in [P_{DG,j}^{min},P_{DG,j}^{max}]\\
Q_{DG,j} \in [Q_{DG,j}^{min},Q_{DG,j}^{max}]\\
\end{cases}, \forall j \in B &&
\label{eq:4}
\end{flalign}
\begin{flalign}
& U_{j} \in [U_{j}^{min},U_{j}^{max}]\quad,\forall U_{j} \in B  &&
\label{eq:5}
\end{flalign}

For accountability of partitioning of the grid into several smaller microgrids, several topological constraints such as clustering constraints, connectivity constraints, etc. are to be considered as mentioned in \cite{b6}. The above constraints are simplified into a sufficient and necessary condition of radiality according to which a radial graph obeys all the above constraints \cite{b34}. A radial graph is characterized by two major properties. Each sub-graph of a graph should be connected and the number of edges should be equal to the number of sub-graphs subtracted from the number of nodes. It has been assumed that each controllable DG could form its own separated island and operate independently to the main grid or in harmony with it.

\begin{figure*}[!htb]
\includegraphics[width=\textwidth,height=0.62\textheight]{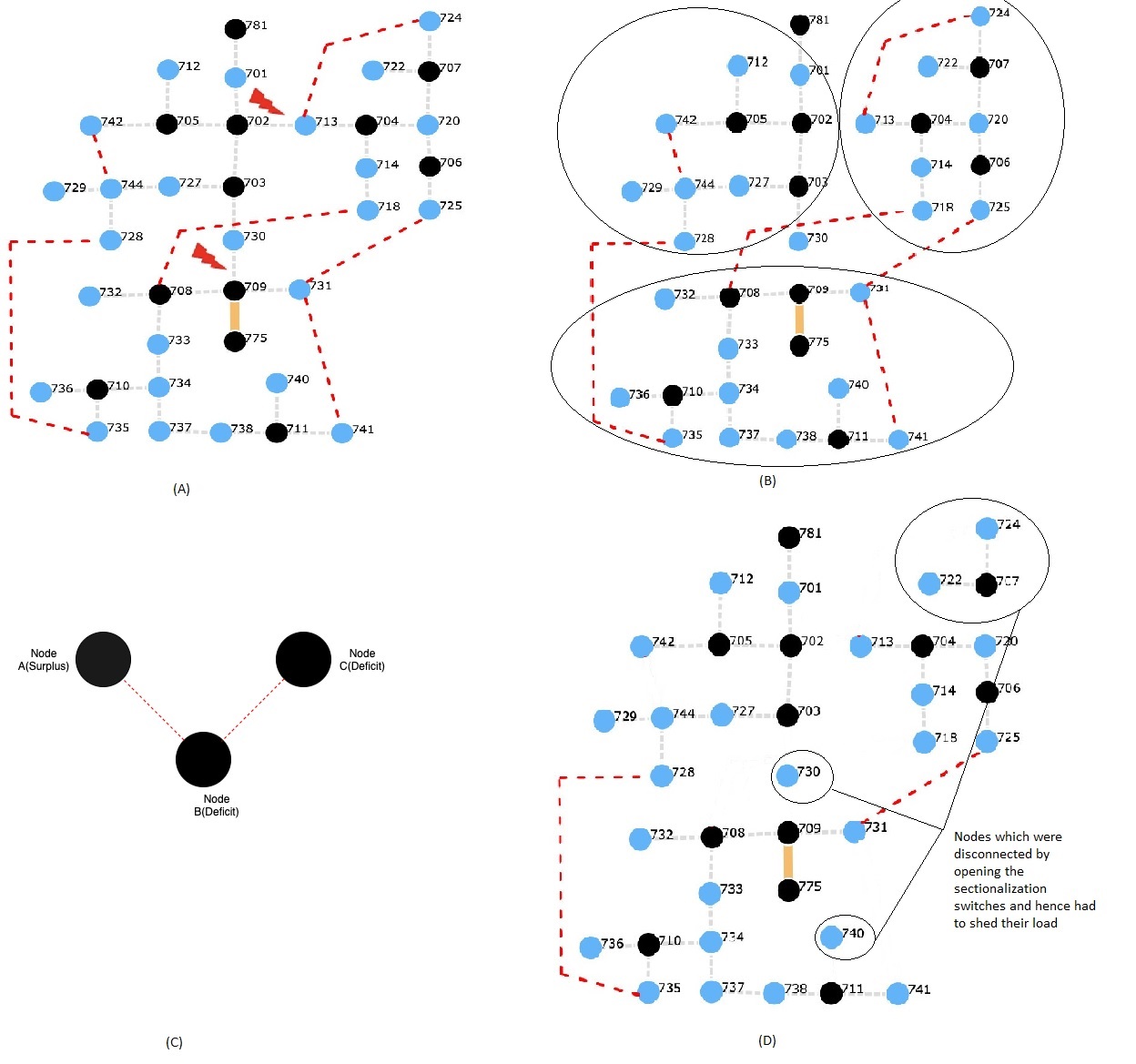}
\caption{A sample grid is shown in (A). The feeders 702-713 and 730-709 are under attack. After the attack, the grid becomes disconnected as shown in (B). It is divided into 3 connected components, some with a power deficit and others with a surplus. The auxiliary graph $G_a$ is formed in (C) as shown. Since this is the optimal set of tie lines to be connected, the nodes whose load needs to be shed are searched for as shown in (D). This is the final grid configuration after islanding and reconfiguration.} \label{fig0}
\end{figure*}

\subsection{Graphical Formulation}
We have modeled the grid as a connected graph $G(V,E)$ consisting of the grid feeders represented as graph edges $E$ and the buses as vertices $V$. In the normal operational mode, the tie lines $T = \{t_1,t_2...t_m\}$ in the grid represented as edges are not connected, all the sectionalization switches are closed and the DGs operate together with the main grid if connected. In the case of a mishap, the tie lines to be closed are identified along with the sectionalization switches to be closed. The DGs in the network could form separate independent islands and operate autonomously. The radial structure of the graph needs to be maintained across all situations to satisfy the topological constraints. The operational constraints are taken into account while construction of the new graph $G'(V',E')$. In the final representation, each unsupplied island needs to shed its load, whereas a supplied island can continue its operation through a DG or the main generator. The computation required to find this newly computed graph is essentially a combinatorial problem, but we aim to solve this in polynomial time through several optimizations in the later sections.

The graph $G$ can be simplified for processing post-attack by modeling a connected component $C_G$ of the graph as a node in a new auxiliary graph $G_a(V_a,T_a)$. The nodes in $G_a$ would either have a surplus or a deficit of power supply. The nodes in $G_a$ which are initially disconnected could be connected by closing the tie lines $T$. The tie lines need to be remodeled for the auxiliary graph by using the mapping for nodes in the original graph $G$ to the nodes in $G_a$. The auxiliary graph $G_a$ can be remodeled back to the graph $G$ for computation of $G'$ by reverse mapping the nodes.
\subsection{Estimation of the minimum load shed}
The auxiliary graph $G_a$ is used for computing approximately the minimum amount of load to be shed. The nodes in the auxiliary graph are initially isolated but can be connected by the tie lines $T$ in the smart grid. To estimate the minimum amount of load shed, we propose to connect all the tie lines across all the nodes, since this would ensure that the maximum number of nodes in $G_a$, hence the maximum nodes in $G$, is interconnected. Hence, the graph $G_a$ can be represented in terms of $G$ as:
\begin{equation}
    G_a(V_a,T_a) = G(V,E\cup T)
\end{equation}
This ensures that the surplus load in certain nodes to be distributed to the nodes in deficit. In case two nodes in deficit are connected, it just adds an extra redundant feeder to the grid with zero power flow. On connecting the tie lines, when two nodes are connected their combined deficit or surplus power will be determined by the addition of the two individual power flow values. This is essentially true since it is consistent with the power flow \cref{eq:1} of the linearized DistFlow model used for the mathematical formulation. It is also powered by the assumption made above that two DGs or even a DG and the main generator can work in harmony with each other. This obtained value for the minimum load shed is used as a reference for the next stages to obtain an optimal configuration for achieving this minimum.

\subsection{Obtaining the optimal islanding configuration}
The optimal islanding configuration can be obtained using auxiliary graph $G_a$ constructed in the previous step and iteratively disconnecting the tie lines which are redundant. For identification of the maximum number of tie lines to be disconnected, keeping the minimum load shed constant, we first order the tie lines in $T$. Now, we remove each tie line from the auxiliary graph $G_a$ to form graphs $G_{1}^{'}(V_a,T_1^{'}),G_{2}^{'}(V_a,T_2^{'})...G_{m}^{'}(V_a,T_m^{'})$. Mathematically, the edges of any such graph $G_{k}^{'}$ can be represented as:
\begin{equation}
T_{k}^{'} = T_a \ominus \{t_k\}
\end{equation}
We compute the load shed with the above configurations. If any of them achieve the minimum, we save them as a potential solution. Now further we compute graphs $G_{12}^{'}(V_a,T_{12}^{'}),G_{23}^{'}(V_a,T_2^{'})$ by removing the tie line $t_2$ and $t_3$ from $G_1^{'}$ and $G_2^{'}$ respectively. More generally, we compute the graph $G_{k_{p_{1}}k_{p_{2}}}^{'}$ with edges:
\begin{equation}
   T_{k_{p_{1}}k_{p_{2}}}^{'} = T_{k_{p_{1}}}^{'} \ominus \{t_{k_{p_{2}}}\}, p_2 > p_1
\end{equation}
The process continues until all the tie lines are considered for a particular branch. Hence, the change in edges could be generalized as: 
\begin{multline}
   T_{k_{p_{1}}...k_{p_{m-1}}k_{p_{m}}}^{'} = T_{k_{p_1}...k_{p_{m-1}}}^{'} \ominus \{t_{k_{p_{m}}}\},\\ p_m > p_{m-1} > ... > p_1
\end{multline}
Note that, we do not consider the configuration where $p_{m} < p_{m-1}$, i.e. removing $t_{p_{m-1}}$ after removing $t_{p_{m}}$, since the final configuration would be the same as that of the removal of $t_{p_{m-1}}$ before $t_{p_{m}}$. It can be said that the removal of tie lines is independent of order of removal. This particularly saves a lot of computations and brings the complexity to polynomial time.

The complexity of the above algorithm can be easily computed by noting the number of trees formed by the above process. It can be observed that the number of trees $N_{T_{k}^{'}}$ formed by removal of at least $t_{k}$ from $T_{a}$ could be given by 
\begin{equation}
   N_{T_{k}^{'}} = 1 + \sum_{i=k+1}^m N_{T_{i}^{'}} 
\end{equation}
This relation can be used to form a recursive relation on the total number of trees formed:
\begin{align}
   \sum_{k=1}^m N_{T_{k}^{'}} = 1(m) + 2(m-1) + 3(m-2)...\sim m^3
\end{align}
This term is of order $\mathcal{O}(m^3)$ where $m$ is the number of tie lines in the grid. Moreover, we need to check for the validity of the constraints for each tree. This would require an additional $\mathcal{O}(n)$ where $n$ is the number of buses. Hence, the overall complexity of the process is $\mathcal{O}(m^{3}n)$.

Finally, the potential configuration with the maximum number of removals is selected as the optimal configuration $G_o$. If there is more than one such combination, then all are considered as potential solutions. This configuration $G_o$ is further adjusted to recognize the nodes whose load needs to be shed.
\begin{figure}[!b]
\includegraphics[width=0.5\textwidth,height=0.55\textheight]{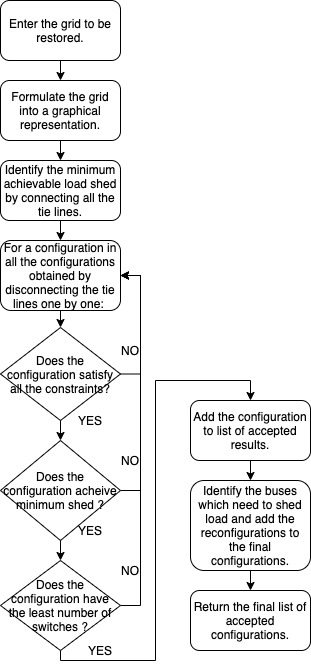}
\caption{The figure gives a brief representation of the algorithm presented in the paper.} \label{fig1}
\end{figure}
\subsection{Identifying the load shed buses}
The configuration $G_o$ obtained in the previous needs to be modified in order to obey all the operational constraints. The topological constraints established aren't violated in the previous steps. The islands with a power deficit need to shed some load in order to satisfy the power balance constraints in \cref{eq:1}. Now if the grid has manual control over the individual buses, it can easily unauthenticated any of the buses it deems fit to shed load, but this is not often the case. Buses for load shed are selected from the terminals of each component with a deficit. Now, each bus might have a weight associated with it. The impact of load shed on it being represented as : 
\begin{equation}
    P_{impact,j} = \varpi_{j}P_{shed,j}
\label{eq:10}
\end{equation}

We consider a connected component $C_{G_o}$ with a power deficit of the configuration $G_o$ and start a multi-source breadth-first search with the terminal nodes as the sources. We identify the nodes which would lead to minimal overall impact according to the \cref{eq:10} but could shed enough load so as to satisfy the power balance \cref{eq:1}. The node might belong to any of the branches but nodes selected in a branch should be continuous and should originate from the terminals. Once these nodes are identified, the sectionalization switches which need to be opened which can disconnect the selected nodes from the connected component $C_{G_o}$. For a $C_{G_o}$ without a DG or the main turbine, all the buses in it need to shed their load in order to satisfy the operational constraints. Moreover, any such line which causes the radial topology to be disrupted such as forming a cycle is disconnected. This step introduces an extra computational complexity of order $\mathcal{O}(n)$ but its negligible in comparison to the previous term which was of order $\mathcal{O}(m^3n)$ 
\begin{figure*}[!htb]
\includegraphics[width=\textwidth, height=0.65\textheight]{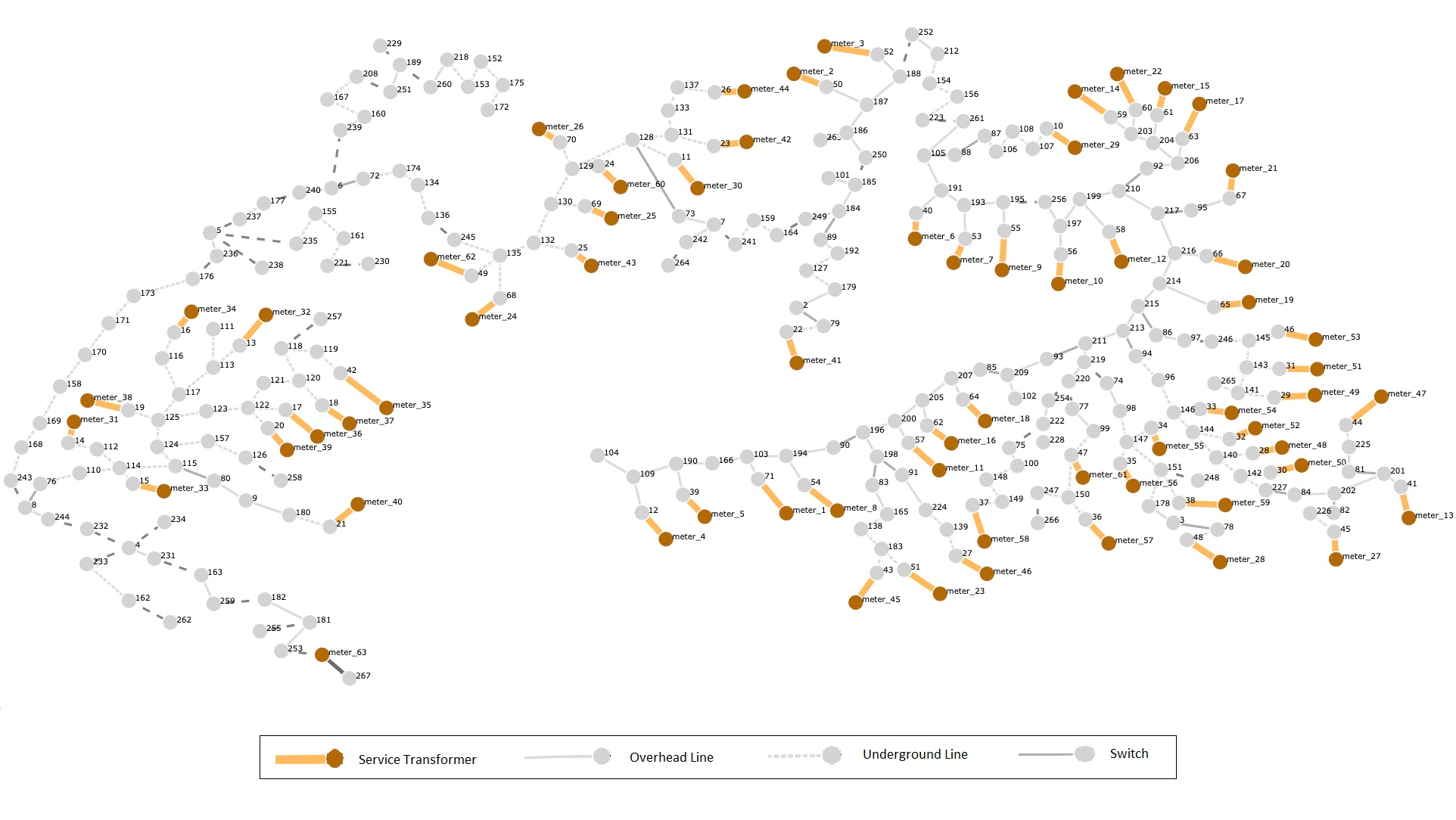}
\caption{The figure shows a simplified representation of the 1069-node grid system. This configuration is used with four taxonomy feeders and interconnected by seven tie lines. Moreover, DGs are placed randomly in the grid.} \label{fig2}
\end{figure*}
Once we perform this for all such connected components, we have our tie lines to be connected along with the sectionalization switches to be opened. Hence, we have all the switching operations which lead to a configuration with a minimum load shed and a minimum number of switching operations, and one which satisfies all the constraints. A brief summary of the complete process is given in \cref{fig0}.

\section {Experimentation and Results}

\begin{figure}[!htb]
\includegraphics[width=0.5\textwidth]{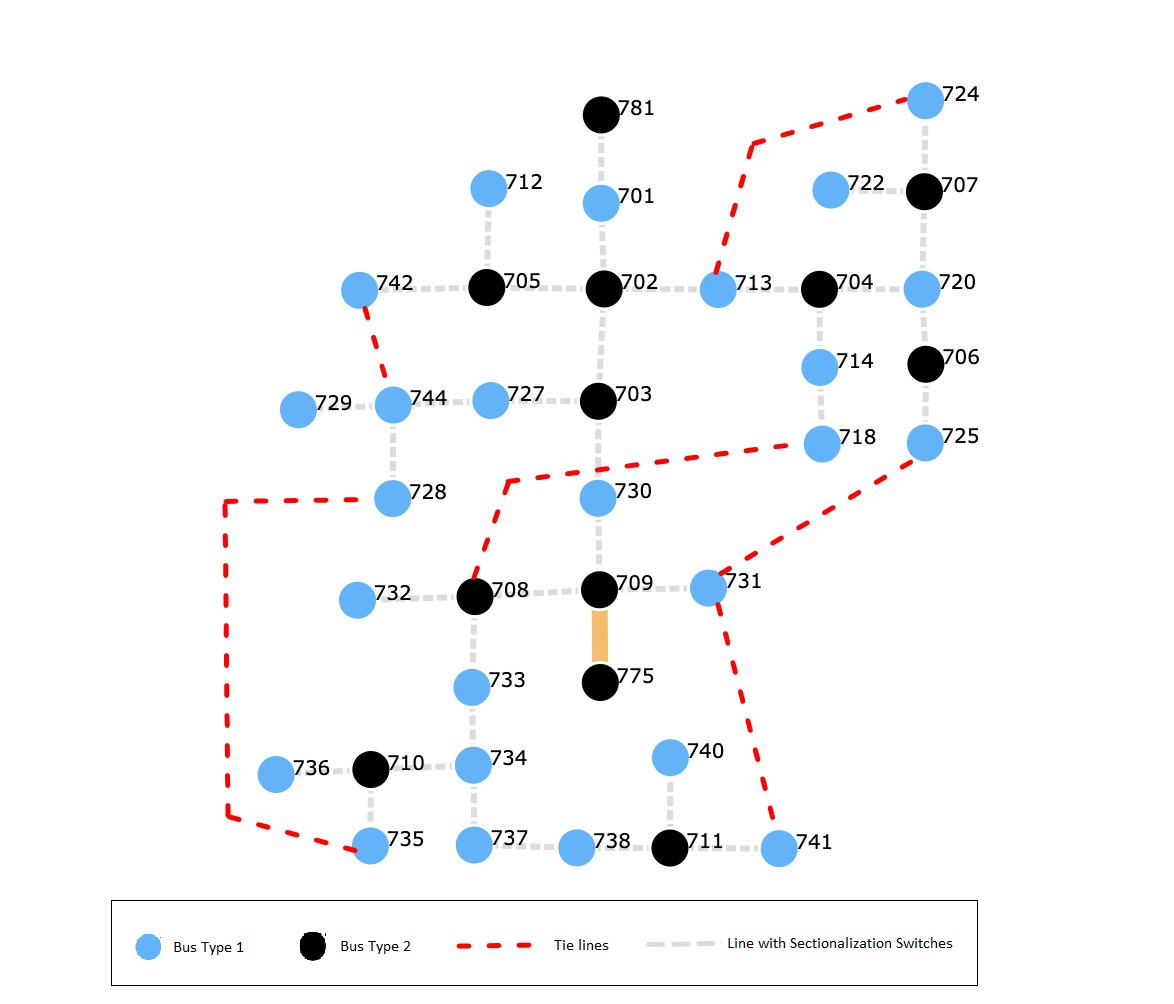}
\caption{The figure shows a simplified representation of the IEEE-37 node system. The configuration is modified by introducing tie lines in the existing grid.} \label{fig1}
\end{figure}

\begin{table*}[ht]
\caption{This table presents the comparison results on the IEEE-37 node system. The time taken for obtaining the solution and the sequence of operations is presented.}
\centering \scalebox{1.0}{
\begin{tabular}{|c|c|c|c|c|c|}
\hline
Scenario & Fault location & \multicolumn{2}{|c|}{ \bfseries Proposed Methodology}& \multicolumn{2}{|c|}{\bfseries Spanning Tree Search}\\
\cline{3-6}
& & Time(in s) &  Switching Sequence &  Time(in s) &  Switching Sequence\\
\hline
1 & 713-704 & 0.081 & Close: 713-724 & 1.335 & Close: 713-724\\
\hline
2 & 730-709 & 0.103 & Close: 708-718 & 0.825 & Close: 708-718\\
\hline
3 & 730-709, & 1.523 & Close: 708-718, 735-728; & -- & --\\
&702-713&&Open: 720-707, 703-730, 711-740&&\\
\hline
\end{tabular}
}
\label{tab1}
\end{table*}

\begin{table*}[ht]
\caption{This table presents the comparison results on the 1069-node R3-12.47-2 system. The time taken for obtaining the solution and the sequence of operations is presented.}
\centering \scalebox{1.0}{
\begin{tabular}{|c|c|c|c|c|c|}
\hline
Scenario & Fault location & \multicolumn{2}{|c|}{ \bfseries Proposed Methodology} & \multicolumn{2}{|c|}{\bfseries MILP based 2-stage restoration}\\
\cline{3-6}
& & Time(in s) &  Switching Sequence &  Time(in s) &  Switching Sequence\\
\hline
1 & 241-159($F_{d}$) & 1.676 & Close: 252($F_{d}$)-266($F_{c}$), 254($F_{d}$)-248($F_{a}$) & -- & Close: 254($F_{d}$)-248($F_{a}$), 252($F_{d}$)-266($F_{c}$) \\
&&&&&75($F_{b}$)-252($F_{c}$), 236($F_{b}$)-256($F_{c}$)\\
&&&Open: 256-254($F_{c}$), 256-261($F_{d}$)&&Open: 220-254($F_{d}$), 136-245($F_{b}$) \\
&&&&&195-256($F_{c}$)\\
\hline
2 & 181-182($F_{c}$) & 8.76 & Close: 252($F_{d}$)-266($F_{c}$), 254($F_{d}$)-248($F_{a}$) & -- & Close: 254($F_{d}$)-248($F_{a}$), 252($F_{d}$)-266($F_{c}$) \\
&&&236($F_{b}$)-256($F_{c}$),244($F_{c}$)-257($F_{d}$) &&261($F_{a}$)-263($F_{b}$), 244($F_{c}$)-257($F_{d}$)\\
&&&&&236($F_{b}$)-256($F_{c}$), 252($F_{c}$)-75($F_{b}$)\\
&&&Open: 256-77($F_{c}$), 256-261($F_{c}$)&&Open: 77-220($F_{c}$), 177-240($F_{b}$) \\
&&&256-261($F_{d}$)&&195-256($F_{b}$), 208-251($F_{c}$) \\
&&&&&220-254($F_{d}$)\\
\hline
3 & 193-195($F_{b}$) & 1.577 & Close: 75($F_{b}$)-252($F_{c}$), 254($F_{d}$)-248($F_{a}$) & -- & Close: 254($F_{d}$)-248($F_{a}$), 75($F_{b}$)-252($F_{c}$) \\
&105-191($F_{d}$)&&&&252($F_{d}$)-266($F_{c}$)\\
&&&&&Open: 195-256($F_{c}$)\\
\hline
\end{tabular}
}
\label{tab2}
\end{table*}

\begin{table}[ht]
\caption{This table presents the percentage reduction in load shed before and after restoration for both cases.}
\centering \scalebox{1.0}{
\begin{tabular}{|c|c|c|}
\hline
Number of Faults & IEEE-37 node & R3-12.47-2 1069 node\\
\hline
1 & 99.45\% & 97.32\%\\
\hline
2 & 63.02\% & 54.48\% \\
\hline
3 & 46.73\% & 35.50\%\\
\hline
\end{tabular}
}
\label{tab3}
\end{table}

The proposed methodology was tested against modified versions of the standard IEEE-37 and IEEE-1069 bus test systems. The reduction in load shed on reconfiguration and the reconfiguration strategy for sample attacks are recorded and compared with results from currently established systems. Each system was imported from a GridLAB-D model with load models of constant power, current and impedance. The algorithm was implemented and tested on MATLAB and GridLAB-D on a system with Intel Core i7 8th generation processor and 16GB RAM.

Each system was imported from a GridLAB-D model with load models of constant power, current and impedence. The algorithm was implemented and tested on MATLAB and GridLAB-D on a system with Intel Core i7 8th gen processor with 16GB ram.

The modified IEEE-37 bus system \cref{fig1} is implemented as shown in \cite{b24} and \cite{b25}. Additionally, the system is augmented with 6 normally open tie lines and 4 DGs at random positions. All the branches are supported by sectionalization switches. The performance of the system is compared with the methods in \cite{b24}. The faults were simulated as given in \cite{b24} so as to be able to easily compare our methods. One advantage of our proposed algorithm is that it works in case of multiple attacks whereas the one in \cite{b24} doesn't. The simulation results are presented in \cref{tab1}. 

In Scenario 1, the attack is on the line connecting buses 713 and 704. As a result of this, the buses to the right of 714 in \cref{fig1} are disconnected from the main grid. These could be connected by joining the tie lines 713-724, 731-725, or 708-718. Some sectionalization switches might need to be opened in case excess load needs to be shed. The best solution to this would be closing the tie line 713-724. This way no-load needs to be shed and the normal working of the grid would be restored in minimum operations.

In Scenario 2, the attack is on the line connecting buses 730 and 709. As a result of this, the buses to the bottom of 709 in \cref{fig1} are disconnected from the main grid. These could be connected by joining the tie lines 735-728, 731-725, or 708-718. The best solution to this would be closing the tie line 708-718. On connecting the line 735-728, we find that an over-current flow through the bus 744 and hence it violates the power constraints. No load would be shed in this scenario.

In Scenario 3, there are multiple attacks on the lines 730-709 and 702-713. As a result, the complete grid is divided into three islands two of which are not connected to any generator. For load restoration, the tie lines 708-718 and 735-728 are connected. This leads to overload in the grid and some load needs to be shed. On computation according to the proposed methodology, nodes 730, 740, 722, 707, and 724 need to shed their load. Hence, the switches in lines 720-707, 703-730, and 711-740 need to be opened. The net load shed was approximately 2.10 MW.

The R3-12.47-2 1069 bus test system \cref{fig2} consists of four taxonomy feeders and numerous tie lines as used by \cite{b23} and \cite{b24}. The model provides a good representation of a real-world populated area. Replicas of a single feeder is made to construct the four feeder system ($F_{a},F_{b},F_{c},F_{d}$) where the feeders are interconnected with tie switches. Hence, with net 1069 buses, 152 branches with sectionalization switches, and 122,586 possible number of radial topologies, the grid is a typical representation of an urban grid. Moreover, the grid is augmented with several DGs to support the primary generators in case of outrages. The system is similarly simulated against attacks and reconfigured by the algorithm suggested in this paper and by \cite{b23}. The data about the time taken for each solution is not available in \cite{b23} but since it uses MILP which is NP-Hard it would be much more time-intensive as compared to our methodology. The simulation results are presented in \cref{tab2}.

In Scenario 1, the attack was simulated on line 241-159 of supply feeder $F_{d}$. As the result, the part of the grid supplied by $F_{d}$ after node 159 was blacked out. For restoration, the tie line between 252($F_{d}$) to 266($F_{c}$) and 254($F_{d}$) to 248($F_{a}$) was connected. This leads to overload in certain buses in $F_{c}$ and hence switches between nodes 256-254 of $F_{c}$ and 256-261 of $F_{d}$ were opened. This restores the load in the minimum number of operations. The net load shed before islanding was approximately 3.5MW but it was restored completely.

\begin{figure}
\includegraphics[width=0.5\textwidth]{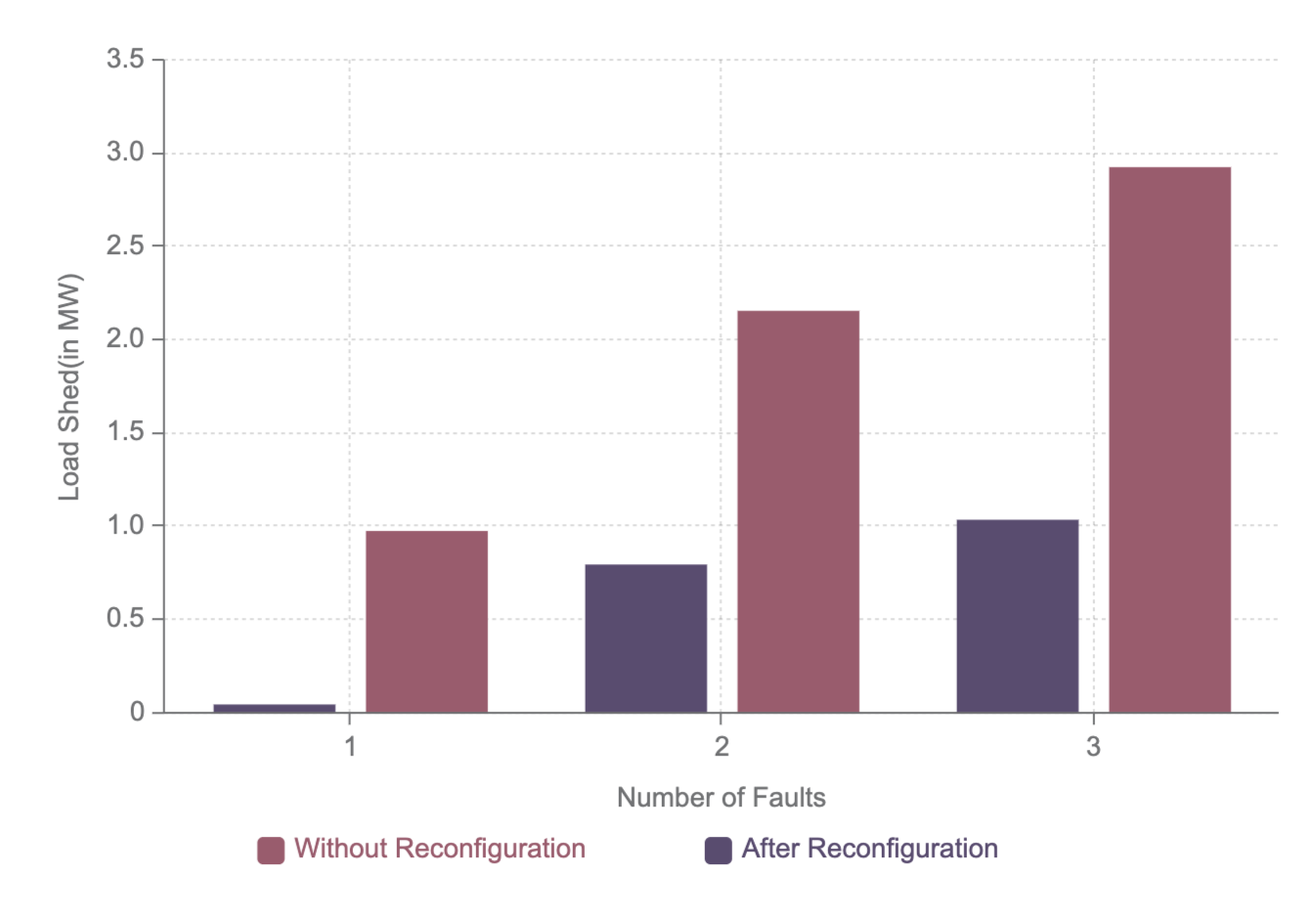}
\caption{This graph presents the load shed pre and post restoration as given by \cref{tab3} for the IEEE-37 node system. The horizontal axis represents the number of attacks whereas the vertical axis corresponds to the amount of load shed.} \label{gr1}
\end{figure}

\begin{figure}
\includegraphics[width=0.5\textwidth]{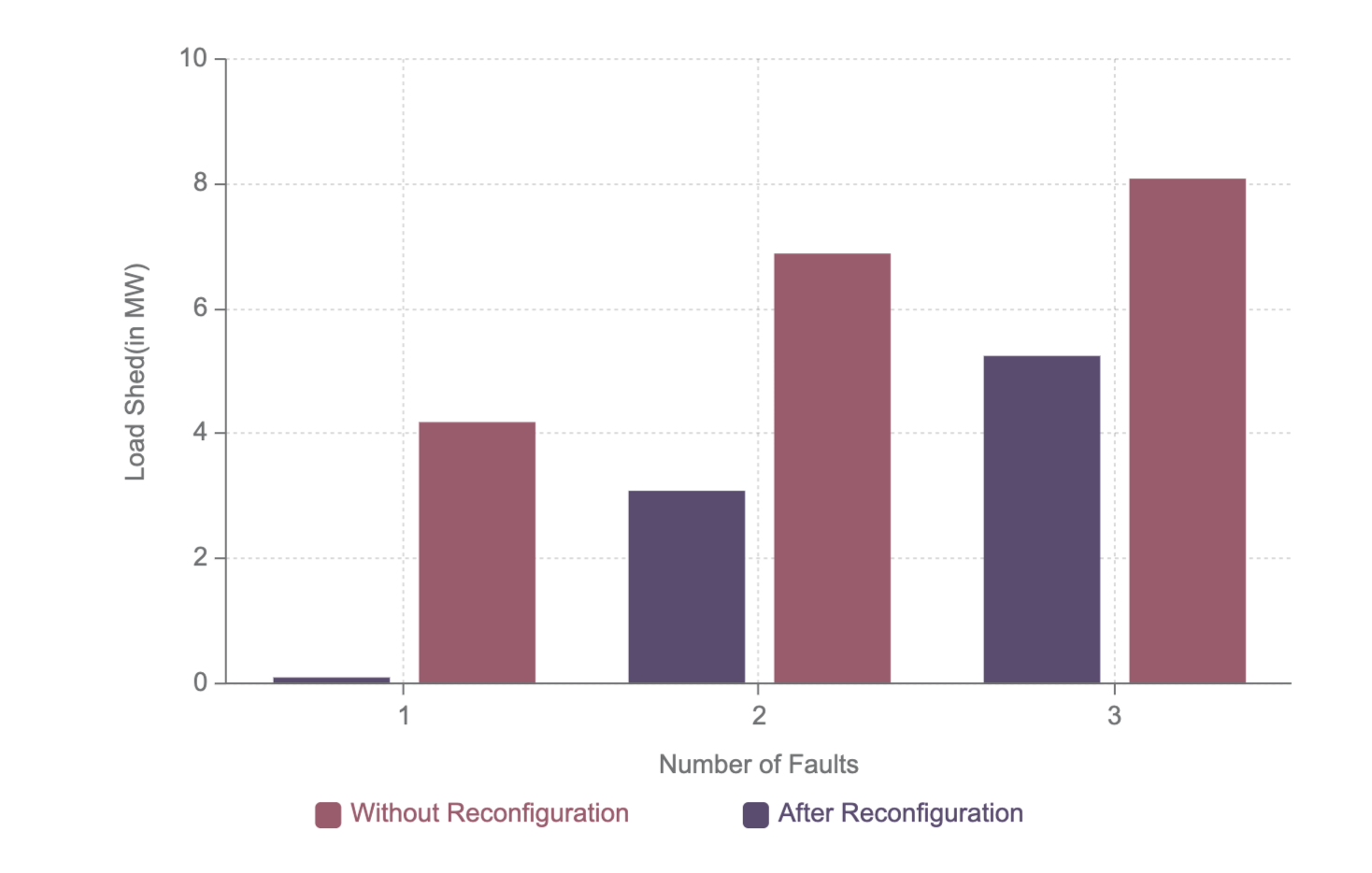}
\caption{This graph presents the load shed pre and post restoration as given by \cref{tab3} for the 1069 node system. The horizontal axis represents the number of attacks whereas the vertical axis corresponds to the amount of load shed.} \label{gr2}
\end{figure}

In Scenario 2, attack was simulated on line 181-182 of supply feeder $F_{c}$. As a result the part of the grid supplied by $F_{c}$ after node 181 was blacked out. For restoration, the tie line between 257($F_{d}$) to 244($F_{c}$), 252($F_{d}$) to 266($F_{c}$) and 236($F_{b}$) to 256($F_{c}$) and 254($F_{d}$) to 248($F_{a}$) were connected. This leads to overload in certain buses in $F_{b}$ and violation of radial topology and hence switches between nodes 256-261 of $F_{c}$, 256-77 of $F_{c}$ and 256-261 of $F_{d}$ were opened. This restores the load in a minimum number of operations.

In Scenario 3, multiple attacks were simulated on line 193-195 of supply feeder $F_{b}$ and 105-191 of supply feeder $F_{d}$. As a result the part of the grid supplied by $F_{b}$ after node 195 and the grid supplied by $F_{d}$ after node 191 was blacked out. For restoration, the tie line between 75($F_{b}$) to 252($F_{c}$) and 254($F_{d}$) to 248($F_{a}$) were connected. This restores the load in a minimum number of operations. The net load shed was 6.5 MW but was completely restored.

The load shed comparison pre and post-restoration is given by \cref{tab3}. Graphical representation of the same is given by \cref{gr1} and \cref{gr2}. for the IEEE-37 node system and the 1069-node system respectively.

\section{Conclusion}
In this paper, we present a novel graph-based approach to the restoration of a distributed smart grid with DGs post-attack. The method considers all the operational and topological constraints and provides the minimum load shed along with the sequence of switching operations to achieve such a configuration. The method was tested against standard IEEE grids and other taxonomy grids which prove that it compares well to the existing methods and provides a result with much lesser switching operations in much lesser time. 

Further improvements may revolve around providing a resilient hardening strategy based on similar principles which would enable the system to incur even lower losses in cases of attacks. Moreover, the algorithm could be modified to account for dynamic changes in grids, isolation of the attack nodes from the rest of the grid in case it has the potential to compromise other nodes, etc. Thoughts could be given on predicting the cost for reconfiguration before attacks.

\end{document}